# Carbon-based Materials as Key-enabler for "More Than Moore"

Franz Kreupl

SanDisk Corporation, 601 McCarthy Boulevard, Milpitas, CA 95035, USA

## ABSTRACT

Carbon-based materials like nanotubes and graphene are heavily investigated as future CMOS-like devices and in interconnect applications. While much of the interest has been devoted to the device aspects in competition to conventional CMOS transistors, the paper here will focus on some less known applications of carbon. Proposed and demonstrated are interconnect applications like highly conductive electrodes for capacitors, gate material or through-silicon vias (TSV), novel non-volatile memories, carbon-silicon Schottky diodes or superior sensors.

## INTRODUCTION

A plethora of carbon allotropes like diamond, fullerenes, carbon nanotubes and graphene has been investigated for application in microelectronics over the recent years. For application in interconnects, nanotubes and graphene has been considered intensively, due to their low specific resistivity. However, for real applications many nanotubes and graphene layers needs to be bundled for parallel operation in order to make their overall resistance comparable or even better than conventional metallization schemes [1].

Graphene is a single planar sheet of $sp^2$ - bonded carbon atoms and graphite consists entirely of individual graphene layers, which are stacked on each other. During this stacking process the specific resistivity of the stacked graphene layers gets worse again and approaches the value of graphite, which in-plane resistivity can be as low as 50 µΩcm and which can be engineered down to 1 µΩcm by doping and intercalation. Only recently, de Heer's group was able to grow graphene stacks which show no deterioration as the layer thickness growths, but it requires a special SiC substrate and very high temperatures of 1550 C [2]. For applications in silicon microelectronics theses conditions are not viable because they would exceed the melting temperature of the silicon substrate.

On the other hand, pyrolytic deposition of carbon layers by CVD is available at temperatures ranging between 350-1200 C depending on precursor gas and deposition conditions. The physical properties of these carbon layers depends heavily on deposition temperature, carrier gas and residence time of the gas species. While the overall electrical properties of these carbon films cannot directly compete with the idealized performance of pure graphene or highly-oriented graphite, optimum conditions can be found to achieve a benefit over competing materials (like highly doped poly-Si, TiN, W or silicides) in terms of temperature budget, resistivity, stress and ease of integration.

Besides these property-based arguments for carbon-based interconnects there is also the aspect of sustainability which needs to be mentioned in this context. Most metals have recently experienced considerable price hikes due to world-wide increased consumption and limited supply. The vision of a carbon-based metallization scheme - maybe not only for microelectronics

- may result in the creation of a technology which can rely on a sustainable supply of abundantly available carbon.

The mere electrical properties of the carbon layer are as important as their interaction with different interfacing materials like high-k materials, semiconductors and metals if it comes to integration on a wafer. Some important questions which need to be addressed are, how the properties of the deposited carbon layer are? Is there diffusion of carbon or from the interfacing materials? How does this impair breakdown behaviors and reliability in insulating materials? What are the effective work-functions for different interface materials? The answers to these questions will lead us to applications of carbon electrode layers in MIM capacitor structures, DRAM capacitors and mid-gap gate-material, attractive for low power CMOS. The question on how these carbon layers interface with silicon will give us in the end powerful low-barrier Schottky diodes which can deal with the high temperature requirements of a front-end process. Therefore, for the first time, carbon-silicon Schottky-diodes can be incorporated in a CMOS flow. Stress, built during deposition of the many different material layers during a process has serious impact on the integration. Thin layers of carbon can reduce the overall stress and can ease the integration. For through-silicon vias the most economic approach of "via-first" was limited up to now to c-Si or poly-Si approach due to the high temperature requirement. Carbon layers can fill very high aspect ratio vias (up to 400) and offer a much better alternative to Si in terms of resistivity, stress and cost. If modifications of the carbon layers can be achieved at reasonable temperatures, the overall resistivity can be dropped down to 10 μΩcm which makes it already attractive for competition with W and Cu wiring. The study of maximum possible current densities in the carbon layers results in the finding of a new non-volatile memory based on the conductivity of different carbon configurations. This will not only enable cross-point memory architectures but could also be implemented in configuring FPGAs.

Finally, the spin transport properties of carbon may be beneficial to solve the problems of the high current densities in spin-torque magnetic memories and for the ubiquitous GMR sensors where Neel coupling limits the sensitivity.

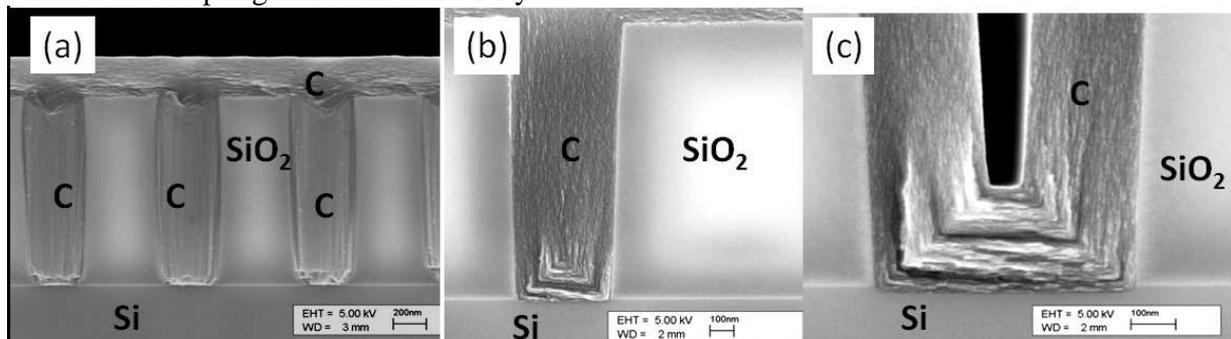

**Figure 1**. (a) Cross-section images of 300 nm wide vias etched in $SiO_2$ which are homogenously filled with the highly conductive carbon layer. (b) Shows a zoomed-in image of a single via where the layered growth of the carbon becomes visible. (c) Illustrates a partially filled via and the cross-section prepared by breaking of the wafer details the layered structure of the carbon filing.

**EXPERIMENT**

The highly conductive carbon layers are deposited by a pyrolytic CVD process with carbon containing gases (acetylene, methane) in a batch process with many wafers in parallel or as individual wafers as described in reference [3,4]. A typical example of the carbon film is given in figure 1 where 300 nm wide vias which have been etched into a $SiO_2$ dielectric layer, have been filled homogeneously with the highly conductive carbon film. The highly conformal deposition is a result of the surface-limited reaction of the deposition process.

In order to investigate the filling properties of the deposition process, very high aspect vias with an aspect ratio of ~400 have been used to demonstrate the ability of this carbon deposition process to fill very high aspect structure features. The vertical holes with a depth of 400 µm and width of 1 µm have been produced by our colleague Volker Lehmann by electrochemical etching of a silicon substrate under appropriate conditions. The pyrolytic carbon used for these experiments was deposited at 950° in a hot wall tube furnace using a methane precursor diluted in hydrogen at a pressure of 300 mbar. The samples were heated up in the hot wall furnace in a hydrogen atmosphere and cooled down in nitrogen ambient flow. The deposition time depends heavily on the "open area", i.e. the surface area which needs to be coated, but the typically deposition rate can be as high as 10 - 60 nm/min.

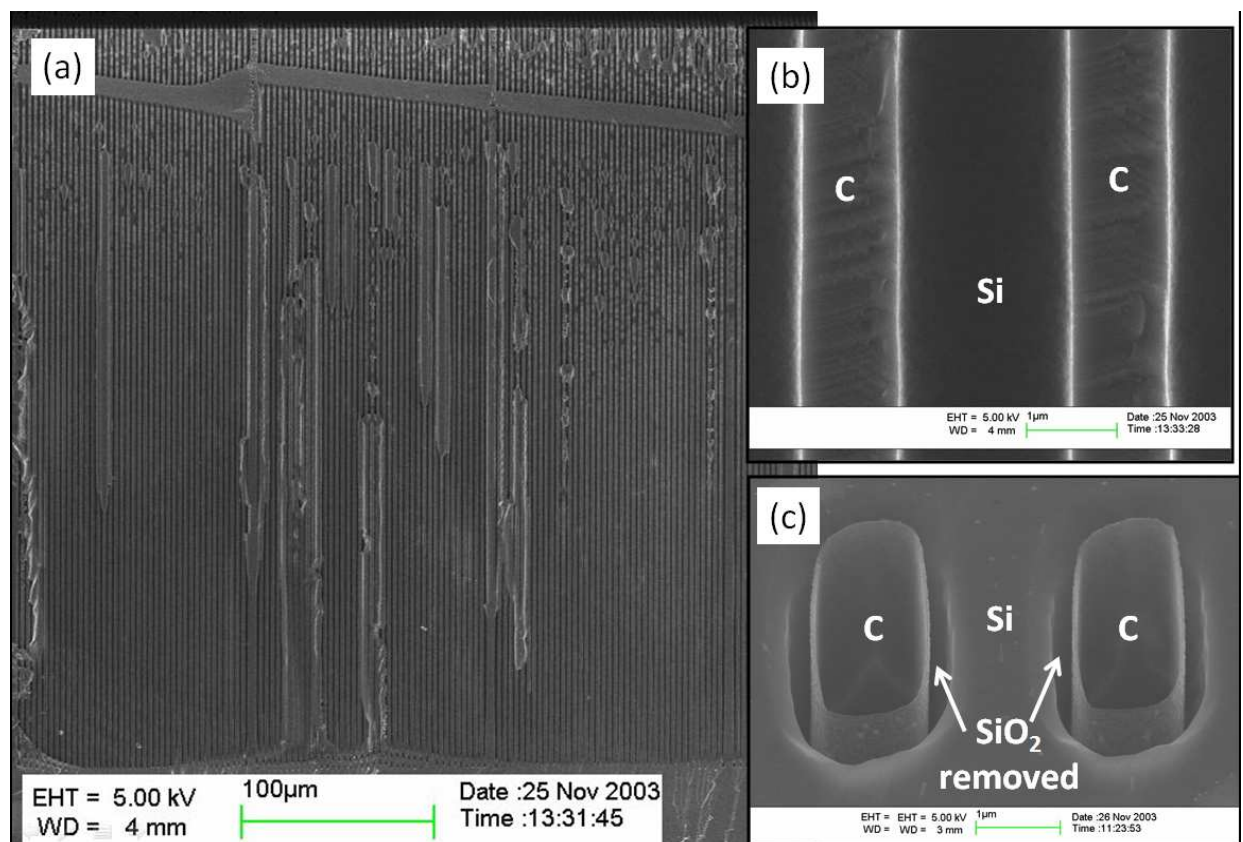

**Figure 2.** (a) Cross-section of a capacitor structure consisting of a Si-substrate with 400 µm deep and 1 µm wide holes with carbon electrode filling. A zoomed-in image is given in (b) and a short etch in KOH separates more detailed the individual components in this carbon-insulator-silicon structure in (c), where part of the silicon and the SiOx layer are recessed and the carbon filling - practically not etched- is protruding from the surface.

As can be observed in figure 2, this deposition process is capable of filling structures with aspect ratios exceeding 400 almost completely within minutes. Figure 2 shows a cross-section of a Si-wafer in which 400 µm deep and 1 µm wide holes have been etched and subsequently filled with the conductive carbon layer. There are a whole bunch of applications which can benefit from this behavior:
- High density metal–insulator–metal (MIM) capacitors which can replace discrete components in analog or RF applications
- DRAM-capacitors in trench and stack applications
- Through-silicon vias (TSV) which can enable 3D package with thermal and electrical vias and flexible backplanes for heterogeneous IC integration
- High-aspect ratio contact holes or vias
- Low contact resistance contact material to silicon and SiC

For the first two applications the introduction of high-k is challenging as many high-k materials are not completely mature for CMOS integration and, depending on the process design, a thermal budget over 1000 C is sometimes required for STI or device activation. As will be shown in this paper, pyrolytic carbon is ideally suited for this application due to its high thermal stability and compatibility to high-k and silicon-based dielectrics even after high thermal budget, good conducting properties, and lower process cost than the more established metal-based metallization schemes. This kind of carbon filling has also wide-ranging applications in TSV-interconnects where the filling depth needs is around 40 – 60 µm, which is roughly the die thickness after back-polishing the wafer back side. During back-side CMP, the carbon pillars serve as excellent CMP etch stop. As carbon withstands easily temperatures up to 1200 C, an economic via-first approach for TSV with metal-like electrical conductivities can be established.

**DRAM and MIM capacitors**

Scaling the DRAM cell to ever smaller technology nodes while maintaining the required 20-30 fF per DRAM cell is a very challenging task and very high aspect structure capacitances needs to be created for both stack- and trench cell concepts. Some melioration can be achieved by the introduction of high-k dielectrics. However, for trench-based DRAM, the introduction of high-k is much more challenging due to the formation of the capacitor prior to device fabrication, which requires STI formation with peak temperatures over1000 C. For the application in trench it is mandatory that the capacitance structure do not deteriorate by high leakage currents, lower break-down voltages or loss of k-value after a thermal stress.

In earlier technology nodes, poly-silicon has been used to fill the high aspect ratio deep trench, but as scaling continues, this is no viable option anymore as the capacitance is reduced by depletion-layer effects in highly doped silicon and the overall resistance of the poly-electrode is not any more acceptable. The carbon electrode allows for a higher electron density during positive bias and yields a capacitance increase of 10-20% due to the missing depletion effect in this metallic-like carbon. Another benefit is a reduction of the leakage current by the higher work-function of the carbon electrode compared to poly-electrode. The work function of the carbon depends on the deposition method and can be adjusted within some range. In the case of lightly nitrogen-doped carbon the work function is around 4.41 eV and gives therefore an improvement of 0.3 V compared to a highly n-doped poly-silicon. The poly-Si fill into the trench is a time consuming process which can last up to 10 hours due to the complicated diffusion process and the high doping which needs to be added. The carbon filling process can fill the trench within 10 minutes – which is a huge productivity advantage [3,4 ]. Figure 3 gives an example of a 6 µm deep trench which is filled with highly conductive carbon within a few minute processing time. Detailed evaluations, as shown in figure 3(b) have shown that the deep trench resistance can compete with TiN metallization. The trench resistance is an important part of the RC-time of the trench capacitor and relates directly to the speed with which the capacitor can be charged. In advanced nodes the poly resistance would lead to 100 kOhm resistance and with TiN or carbon, this value can be brought down to the 20 - 30 kOhm range again.

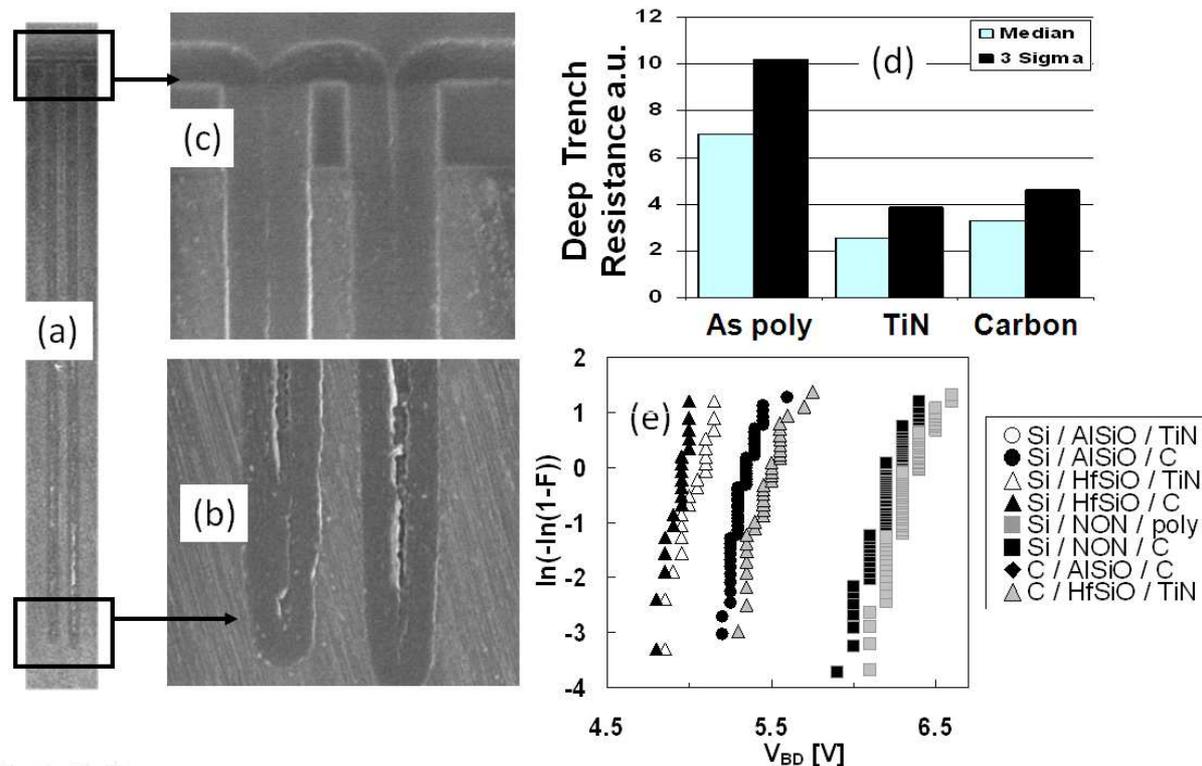

**Figure 3.** Cross-section of a 6 µm deep trench filled with carbon as conducting electrode. A blow-up off the upper part is shown in (c) and the bottom of the trench is show in (d). The deep trench resistance is compared to TiN and highly doped poly-Si in (d) and the reliability distribution of the breakdown of different capacitor electrodes are plotted in (e) [5].

The lower breakdown voltage in the distribution of figure 3(e) in case of carbon electrode compared to poly electrode comes due to a difference in the work function of the two electrodes which causes a shift in the flat-band voltage and this results an increased electric field across the NON dielectric for the case of carbon electrode compared to poly electrode for the same applied voltage. In the case of a MIM structure, the depletion effect is missing on two sides of the high-k dielectric which increases the field even more and causes a further reduction of the breakdown voltage. Although the carbon results in lower breakdown voltages compared to poly, there exists a large margin for the product conditions even with for the carbon-insulator-carbon (CIC) structures to guarantee the 10 years life time. The properties of different approaches of carbon as DRAM electrodes have been already published extensively [5, 6] and have even led to the successful integration into the products based on 58-nm deep trench DRAM technology at Qimonda.

It should be emphasized that the concept of the DRAM capacitors may easily be extended to embedded on-chip capacitors and the compatibility of the carbon electrode with a wide range of high-k materials can enable better capacitors. An example of such a CIC-capacitor is shown in figure 4, where carbon is used on both sides of a high-k material based on AlOx in a short, only 2.6 µm deep trench. The 10 - 17 nm thick carbon layer on both sides of the dielectric is in turn contacted by a highly doped crystal silicon on the outer side and with poly-Si in the trench hole. The trench could of course be filled completely with the more conductive carbon layer, but in order to allow further processing in a production fab, the carbon has been covered with the well known poly-Si. This avoids the intricate contamination protocols, which are usually required if "new materials" are introduced into a semiconductor production line and interfaces the subsequent process steps with the well-established poly-Si surface.

The contact between the crystal silicon and the carbon is achieved by growing carbon on Si. On the inner side of the trench, poly-Si is grown on the carbon surface. The C-Si combination can only be used because the Schottky-barrier of the C-Si contact is reasonable low – a property which we will use to fabricate excellent Schottky diodes – and therefore allows a fast charging and discharging of the capacitor, as required for fast operations and transient filtering. In order to get a low contact resistance at the C-Si interface extreme care must be taken to maintain a very high surface doping of the silicon on one side and omitting the existence of a thin nitride or oxide layer during the preparation and growth of the carbon layer. If poly-Si is grown on carbon, adequate care needs to be taken for the preparation of the carbon surface, because any porous low density carbon surface will lead to the formation of a SiC-like layer which would result in an insulating, high contact resistance. The properties of the carbon surface are mainly determined by the way the growth process is terminated and which kind of gas is used to cool down the wafers after deposition. Any preceding wet-chemical treatment in combination with the atmosphere at the heat-up for poly-Si deposition can induce a thin low density carbon layer which would result in SiC formation and an increased contact resistance. However careful processing allows for a reasonable low contact resistance, so that the silicon-carbon material system can be used.

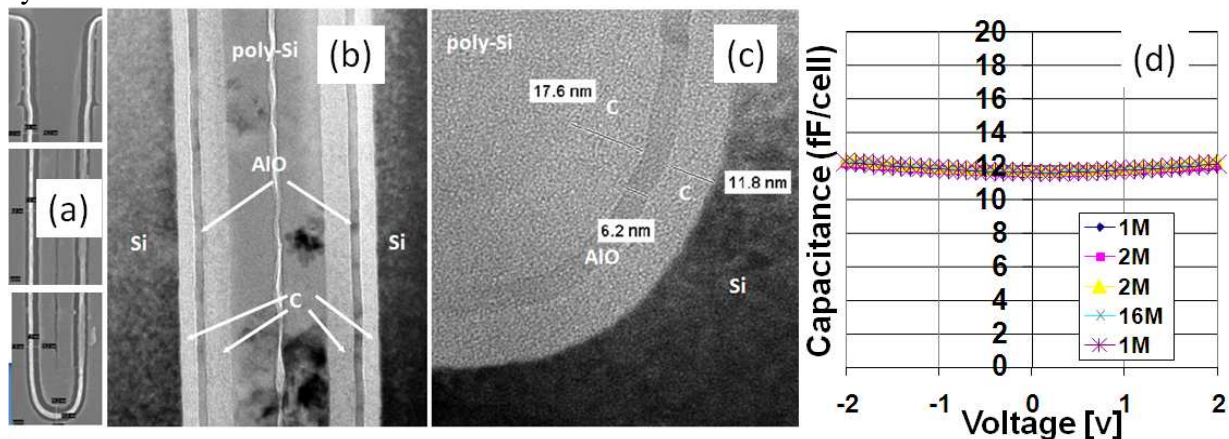

**Figure 4.** Carbon-Insulator-Carbon (CIC) capacitor fabricated with an AlOx-based dielectric. The cross-section of a short, 2.6 µm deep trench is shown in (a) on the top, middle and bottom of the trench. TEM images of the middle portion are shown in (b) and (c) gives a detailed image of the bottom part of the trench CIC. The metal-like, voltage-independent capacitance is plotted for 1M to 16M cells in (d).

The bias voltage dependent capacitance per trench cell for array sizes varying between 1M and 16M is plotted in figure 4(d). It shows nicely the metal-like, voltage-independent capacitor properties. A capacitor density of roughly 300 fF/$\mu m^2$ can be achieved with this design, outpacing the 3fF/$\mu m^2$ obtained for usual SiN MIM capacitors by a factor of 100. However, detailed frequency-dependent measurements of capacitor linearity still need to be done.

**Carbon-Silicon Schottky Diodes**

Schottky diodes have the advantages of low forward voltage drop and fast switching speed. Due to their excellent high frequency performance, they are widely used in filter application for clamping and clipping of supply voltages, reverse polarity protection, in mobile phones as network detectors, as WLAN detectors, in 24GHz radar, RFID and other very high frequency mixer applications. In fact, every frequency multiplier operating at over 100 GHz has a Schottky diode as a central element. In order to increase performance, decrease the supply voltage and packaging costs, it would be advantageous not to have Schottky diodes as separate IC on a board but to be able to integrate Schottky diode into monolithic CMOS design. However, Schottky diodes are fabricated by depositing metals on n-type or p-type semiconductor materials such as Si, GaAs and SiC and these metal–semiconductor contacts, and consequently the Schottky barrier height, are very sensitive to process and the temperature budget. The properties of the formed silicide contacts vary considerable with this process conditions and so would the properties of the diode. The variation is evident even for the standard products in the wide range of specification in the product specification sheet of the different suppliers. Recently, some groups have achieved to make Schottky diodes integrated with capacitors on a CMOS technology but the straight forward integration with analog and logic components have not yet achieved.

Here we will show that a carbon-silicon (C-Si) Schottky diode has outstanding electrical performance and is high temperature stable - a fact which helps to increase the peak current compatibility [4]. Regular metal-based Schottky diodes cannot withstand high peak currents because the metal diffuses into the Si base layer and destroys the diode properties. In order to perform a direct benchmark with a commercially available product we used a BAT17 Schottky diode vehicle where substrate, all layouts, doping levels, guard ring implants has been kept the same as the commercial product, but instead of having the standard metallization for the formation of the Schottky contact, we used the carbon-layer to establish the Schottky contact. A thin metal layer consisting of 5nm Ti and 20 nm Au was deposited on the carbon by a shadow mask and this metal layer served as an etch mask to remove the carbon layer from the open area. Figure 5 (a) gives a top view on the circular diode active area and the TEM image shows the atomically resolved carbon-silicon interface. The forward current of the diode with metal silicide (Bat17) and with carbon are plotted in figure 5 (b). The C-Si diode delivers 15 times more current at a given voltage due to its lower Schottky barrier, which has been evaluated by the Norde-plot to be 0.41 eV for n-doped and 0.59 for p-doped silicon. The ideality factor for the diodes varied between 1.1 and 1.6 and this could be correlated with the presence of an oxide at the interface, which has grown during the heat ramp. At current levels above 1 mA, the missing metallization for the C-Si diode leads to an increased serial resistance caused by current crowding at the small measurement area from the probe tips, whereas the product BAT 17 is packaged and wire-bonded. Interestingly enough, the reverse leakage at low bias is higher for the C-Si diode, but lower than the BAT17 diode at higher reverse voltage. As summarized in figure 5(d), this means that the blocking voltage of the C-Si is 11V instead of only 4 V as for the BAT17 diode at a specified leakage current of 10 µA. The leakage increase at low voltages is a direct consequence of the lower Schottky barrier height and can be easily explained by this. The field at the edge of the diode, where the guard ring is, is much higher for the silicided contact, because the silicidation process consumes a bit of silicon and leads to a very sharp metallic edge feature. The field enhancement at this edge leads to an earlier breakdown at higher voltages. In contrast to this, carbon does not develop this sharp feature at the edge, due to the missing silicidation and as a consequence develops a lower field at the same applied reverse voltage. The steep increase of reverse current at 12 V is the breakdown of the reverse biased pn-junction of the guard ring. The behavior of the diode does not noticeable change even if it is subjected to a 1000 C anneals, which allows the combination with conventional CMOS front-end processes.

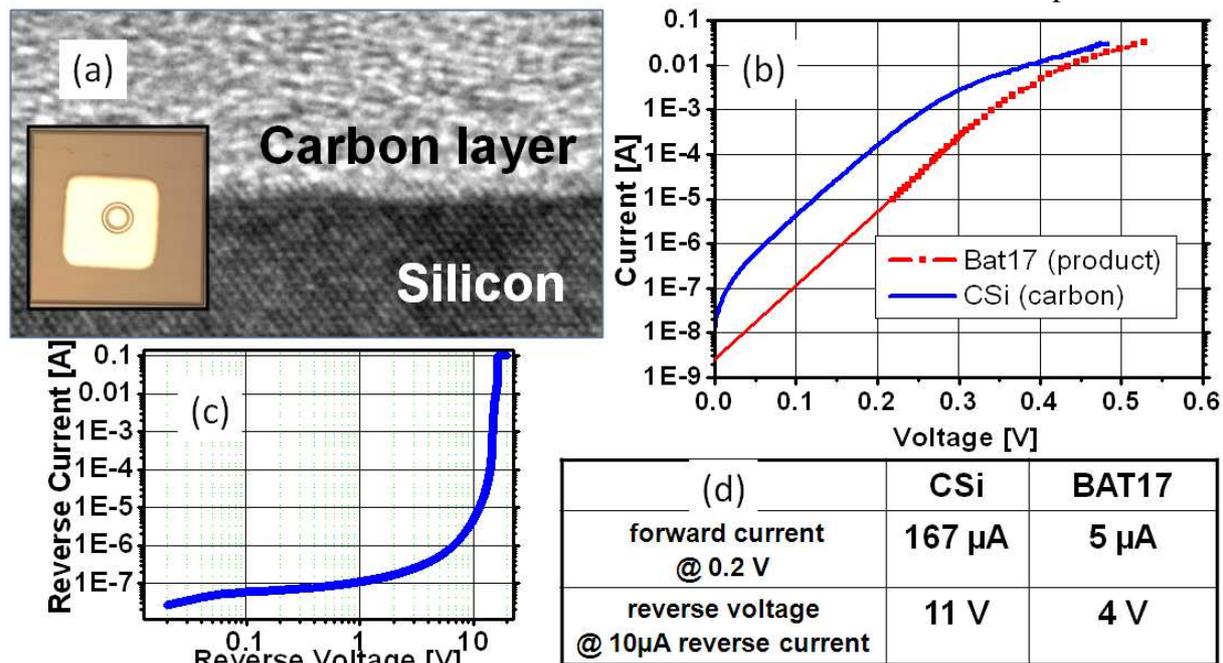

**Figure 5.** Carbon-silicon Schottky diodes have been fabricated on a BAT17 vehicle and benchmarked against the commercially available BAT17 Schottky diode. A TEM cross-section of carbon deposited on silicon is shown in (a) together with a inset of a top-view of the circular diode area with Ti/Au metallization contact on top of a carbon layer. The forward current characteristic of the C-Si diode and the product is plotted in (b). The C-Si gives 15 times more forward current than the regular diode. The reverse characteristic of the C-Si diode and a benchmark table is shown in (c) and (d) respectively.

From the direct comparison of the two Schottky-diodes, it is possible to conclude that the capacitance of the diode which goes roughly proportional with the junction area can be reduced by a factor of 15 while maintaining the same current drivability, if one is using the C-Si contact. This has important consequence in high-speed applications because the switching speed is impacted by this capacitance.

The C-Si interface is capable of withstanding current pulses in the order of 30 MA/cm$^2$ without deterioration, which is very interesting for filter applications. There are also applications in non-volatile resistive memories which can benefit from a diode with high drive current and low voltage drop [7]. As the dynamic energy consumption in a circuit goes quadratically with the voltage, every possible voltage reduction will result in lower power consumption. There are a whole bunch of resistive memory applications like PCRAM, CBRAM, carbon memory or transition oxide memory which are in need of a diode which can deliver peak currents in the order of 30 MA/cm$^2$. If the thickness of the low doped epi-n-layer of the C-Si Schottky diode is reduced to 100nm or below, the serial resistance of the diode can be dropped considerable and results in C-Si diodes with high current pulse capability [7].

In summary, carbon facilitates high performance Schottky-diodes, and due to the high temperature stability of the C-Si interface and the low Schottky-barrier of this interface, the integration in CMOS chips is achievable for the first time.

**Current Carrying Capability, Work-function and Conductivity**

For extended interconnect applications it is important to know the specific conductivity of deposited carbon as well as the maximum current carrying capability of the carbon structures. For applications as gate-electrode, the work-function needs to be evaluated by terraced oxide, C(V) and Fowler-Nordheim-plots in order to determine the flat-band voltage for a specific transistor technology and channel doping.

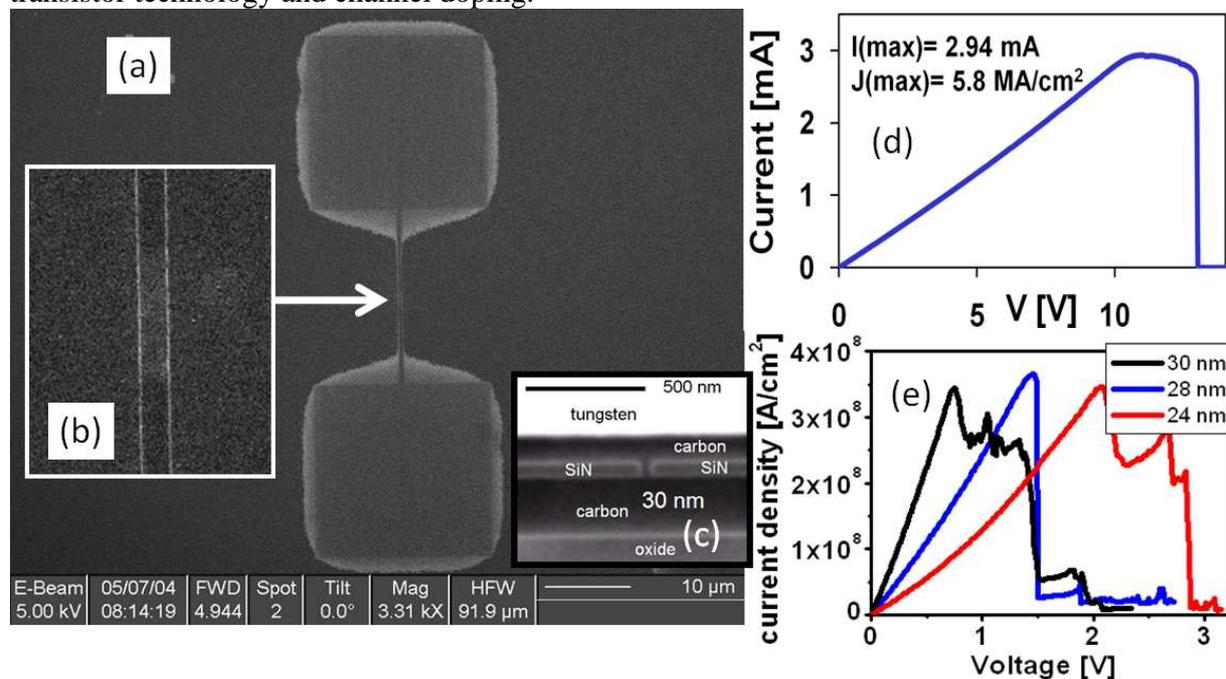

**Figure 6.** Simple interconnect structures like lines and vias have been fabricated to investigate maximum current density and conductivities of the carbon structures. A 13 µm long line with $5.1 \times 10^4$ nm$^2$ cross-section is shown in (a) together with the inset (b) where the damaged region can be seen in the middle of the line after a current of 2.9 mA (d) has passed through the structure. A 30 nm wide via cross-section is shown in (c) together with the current stress curves in (e) for different via widths.

The results of investigations on lines and vias, shown in figure 6 have been recently summarized [8] and a typical specific resistivity ρ in the order of 1-2 mΩcm has been evaluated for a specific carbon deposition in this experiment. A cautious note needs to be added regarding the observed sustaining current densities of carbon. Pyrolytic carbon is used as crucibles for evaporation purposes and it can decompose almost every other material, because the delivered power density, given by $j^2 \cdot \rho$, is very high. The line in figure 6 (a) has been measured in air and although it was covered with a thin oxide layer, it broke down at 5.8 MA/cm$^2$ (see figure 6(d)) because of a solid state reaction of carbon with the oxide and the decomposed oxide layer is permeable for air and the wire burned off as can be seen in figure 6 (b). For vias, shown in figure 6 (c) and (e), measured in vacuum and with a nitride dielectric, the maximum current densities amounts to 360 MA/cm$^2$ at which a phase transition in the carbon towards a more insulating state is supposed to occur.

The work function of a carbon electrode on a 13 nm thick silicon dioxide has been evaluated with C(V)-measurements and analysis of the Fowler-Nordheim-plots of the reliability measurements. The results, shown in figure 7 (a) and (b), gives a work function of 4.59 eV from C(V) and 4.56 eV from the FN-plot. Figure 1(a) shows a well behaved low-frequency C(V) curve with a small amount of interface traps close to the conduction band. This is indicated by the small peak of the high-frequency C(V) curve. Close to threshold voltage, the interface traps contribute to minority carrier generation which increases the capacitance initially. At strong inversion the interface traps are shielded by inversion charge. From this it is concluded that carbon can serve as a mid-gap gate-electrode material, which would allow cost savings and may eliminate some of the problems with NBTI from the boron doping in a traditional CMOS process.

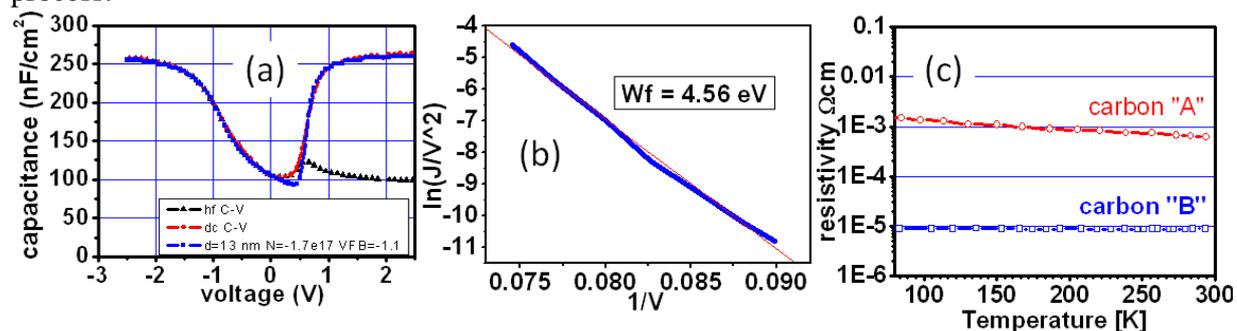

**Figure 7.** The work-function of the carbon-layer is determined on a 0.006 cm$^2$ big CMOS capacitor. The work function is 4.59 eV taken into account the observed interface charge (a). The work function as determined from the Fowler-Nordheim leakage current in reliability measurements is 4.56 eV (b). The conductivity of an un-doped carbon layer "A" shows a negative temperature coefficient with a resistivity below 1 mΩcm at room temperature whereas the resistivity of a doped carbon layer "B" is close to constant for this temperature range and achieves low values in the order of 10 µΩcm.

The work function and conductivity of the carbon layer depends severely on the deposition process and may also be impacted by the exposed substrate surface. Process cost and safety consideration can lead to use nitrogen gas as co-flow, which is cheaper and safer than argon, helium or hydrogen. The use of nitrogen leads to incorporation of large amounts of nitrogen in the carbon-layer. This not only impacts the work function, and leads for instance, for a carbon layer with N-content in the percentage range to a work-function of 4.4 eV but also leads to a more isotropic growth of the carbon layer, which drives the resistivity to higher values. SIMS and NMR analysis revealed in addition hydrogen content of several percentage and even oxygen. The residence time of the gas species seems to play an important role for the layered graphene-like growth of the pyrolytic carbon. As shown in figure 7 (c), the resistivity of un-doped carbon can drop below 1 mΩcm and more elaborate doping schemes can lead to carbon layers with 10 µΩcm specific resistivity. Therefore, un-doped carbon is already superior to highly doped poly-Si and even crystal silicon. High temperature or high current density stress only improves the conductivity of carbon without the perils of doping out-diffusion and doping deactivation as it is the case with doped of silicon. If the resistivity of doped carbon can be brought down to the 10 µΩcm range at reasonable processing temperatures than carbon has a real chance to substitute a number of metals in interconnect applications [3,4,7].

**Carbon Memory**

Carbon seems to be an ideal material for being electrically manipulated in its conductive state. Carbon exists in many forms, the most prominent ones being the $sp^2$-dominated graphitic form with its high conductivity, which was used in the previous parts of the paper and the $sp^3$-dominated diamond form which has a low conductivity. Recently, it has been shown that one can switch from the low conductive state to the highly conductive state and vice versa by applying appropriate current pulses to the carbon element [9,10,11,12]. The switching is illustrated in figure 8 (a) below, where MD calculations have been performed to simulate the switching mechanism [11]. The anneal pulse which is in the order of tens of nanoseconds forms $sp^2$-rich filaments and the very short quench pulse (~1 ns) results in an disordered, $sp^3$-rich quenched state. The two conductance states can be used to represent logical states as seen in figure 8 (b). In contrast to the conventional phase-change materials, carbon is of mono-atomic nature and therefore may be scalable to very small feature sizes (even single bonds).

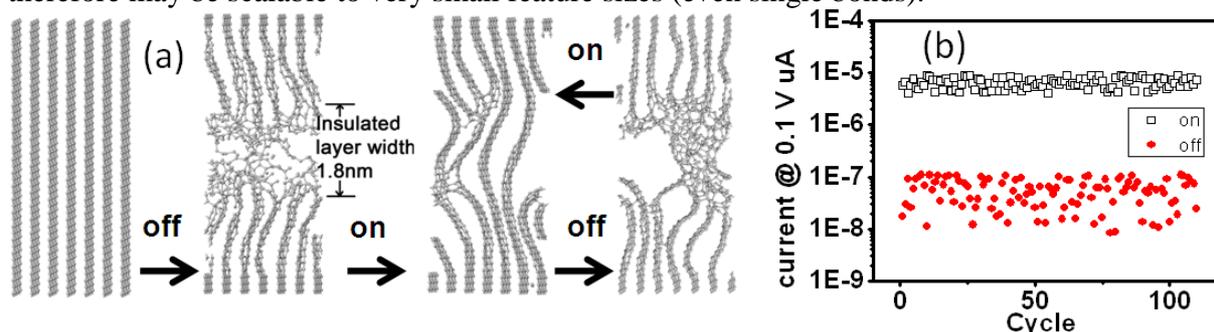

**Figure 8.** The mechanism of the carbon memory is illustrated in the MD-simulation pictures in (a) (Images are by courtesy of Yu He et al. [11]). The on- and off-states of a carbon memory cell, as shown in (b), are manipulated by short voltage pulses to change from a high to low resistance state and vice versa.

**DISCUSSION**

In the preceding examples, many different applications in microelectronics for highly conducting carbon films have been given – some of them have even made it already into real products. The major task here is similar to the in-situ carbon nanotube growth: bring down temperature while maintaining high quality properties, like high conductivity. Unfortunately, there is not much room left to discuss the application of the carbon layer in magnetic sensors and MRAM devices. The unique spin-filtering property of this layer can enable GMR sensors with much higher sensitivities and could even solve the reliability problem of the tunneling barrier of the spin-torque MRAM non-volatile memory cell, as described closer in the references [13,14]. All these attractive features of carbon make it worth to work towards an implementation of carbon as a sustainable interconnect material - just like carbon nanotubes, but even without the need of a precious metal catalyst.

**ACKNOWLEDGMENTS**

Many people at Infineon's Corporate Research Department and at Qimonda have contributed to the results presented here and I especially would like to acknowledge the contributions of Robert Seidel, Gernot Steinlesberger, Werner Pamler and Eugen Unger in the early phase and Andrew P. Graham, Maik Liebau, Georg S. Duesberg, Guenther Aichmayr, Tim Boescke and many more people (see ref. 5,6,9) from the Dresden Qimonda Fab in the later phase of the project as the process has been transferred into the production. I also would like to acknowledge the continuous support of Wolfgang Hoenlein throughout the many phases of this development.